\documentclass[%
 reprint,
 amsmath,amssymb,
 aps,
prl,
]{revtex4-1}

\usepackage{subfigure}
\usepackage{graphicx}
\usepackage{epstopdf}
\usepackage{dcolumn}
\usepackage{bm}

\begin{document}

\preprint{APS/123-QED}

\title{  Generating X-rays with orbital angular momentum in free-electron laser oscillator}

\author{Nanshun Huang$^{1,2}$}


\author{Haixiao Deng$^{3,1}$}%
\email{denghaixiao@zjlab.org.cn}

\affiliation{%
$^1$Shanghai Institute of Applied Physics, Chinese Academy of Sciences, Shanghai 201800, China}%
\affiliation{%
$^2$University of Chinese Academy of Sciences, Beijing 100049, China}%
\affiliation{%
$^3$Shanghai Advanced Research Institute, Chinese Academy of Sciences, Shanghai 201210, China}%


\date{\today}

\begin{abstract}

The generation of light beams carrying orbital angular momentum (OAM) from a free-electron laser at short wavelengths has attracted considerable attention as a key resource in several fields and applications. Herein, we present a facile method to generate intense and coherent OAM beams from an X-ray free-electron laser oscillator (XFELO). We use the Bragg crystal as both a reflector as well as mode selector, in both longitudinal and transverse modes, which enables the specified resonance deviation that maximizes the single-pass gain of the high-order OAM mode of interest. This can allow the amplification and saturation of OAM beams in a typical XFELO configuration. Our results show that 150~$\mu$J fully coherent, hard-X-ray pulses carrying first-order OAM can be generated without utilizing any external elements such as optical mode converters or elliptically polarized undulators. This simple and straightforward method can pave the way for the further customization of the transverse-mode operation of X-ray free-electron laser oscillators.
     

\begin{description}

\item[PACS numbers] 41.60.Cr

\end{description}
\end{abstract}

\maketitle


Over the past few decades, the generation and application of custom light fields with structured intensity, polarization, and phase have attracted significant interest from researchers ~\cite{10.1038/s41566-018-0328-8,Shen2019,Sroor2020}. In particular, light beams with a spatial phase dependence of exp($il\phi$), where $\phi$ denotes the azimuthal coordinate and $l$ an integer referred to as the topological charge (order)~\cite{Allen1992}, carry orbital angular momentum (OAM), which is currently one of the most intensively studied topics in optics. In this regard, OAM beams in the visible- and infrared-wavelength regimes have already been utilized in diverse areas of applications such as micro-manipulation, quantum information, and optical data transmission~\cite{Shen2019}. In the X-ray regime, the use of OAM beams can enable the direct alteration of atomic states through OAM exchange\cite{Sakdinawat2007} and facilitate the development of new methods to study the quadrupolar transitions of materials~\cite{VanVeenendaal2007}. However, the practical applications of X-ray beams with OAM are currently limited owing to the lack of suitable optics and the difficulties of realizing practical coherent light sources.

In general, it is common to generate OAM beams by inserting optical elements~\cite{Peele2002, Vila-Comamala2014} such as programmable spatial light modulators, stepped phase plates, and spiral Fresnel zone plates into the propagation path of light. However, these direct optical manipulation approaches may not be practical or available for application to modern X-ray free-electron lasers (XFELs)~\cite{emma2010first,ishikawa2012compact,milne2017swissfel,Kang2017,Decking2020,Pellegrini2016,Feng.2018}, which are currently the brightest source of X-rays for scientific applications. The XFEL can operate in a high-gain mode typically referred to as self-amplified spontaneous emission and a low-gain mode referred to as the oscillator mode. In this context, researchers have studied and proposed several methods and techniques for generating OAM beams in a high-gain XFEL ~\cite{10.1103/physreva.77.063831,10.1103/physrevlett.102.174801,10.1103/physrevlett.106.164803,Hemsing2013,10.1103/physrevlett.112.203602}. One such practicable technique involves higher-harmonic X-ray emission from a helical undulator~\cite{10.1103/physrevlett.100.124801}. This approach is recommended for application in the afterburner scheme~\cite{Hemsing2013}, in which the helical undulator follows a long section of inversely tapered undulators. These inversely tapered undulators serve as electron bunchers operating at a harmonic resonance frequency of the helical undulators. In the following helical undulators, the bunched electron beam can emit intense harmonic radiation carrying OAM. However, external spectral filters or other techniques are required to separate OAM beams from the mixed radiation because the fundamental radiation is emitted predominantly in the helical undulators as well. Meanwhile, in a high-gain seeded XFEL, a practicable approach for producing OAM beams is to use a helically bunched beam~\cite{10.1103/physrevlett.106.164803}. In this approach, the electron beam is first helically bunched by means of harmonic energy modulation followed by density modulation and is then sent into a long planar undulator operating at the fundamental resonance frequency. However, this scheme may fail at short wavelengths as it requires a harmonic modulation process, for which an intense seed laser is required. 

Meanwhile, X-ray free-electron laser oscillators (XFELOs) are low-gain devices that can produce intense, fully coherent hard X-rays~\cite{Kim2008,Dai2012,Li2018}. In an XFELO, the X-ray pulses circulate in a low-loss optical cavity formed by multiple Bragg-reflecting crystals. In general, XFELO pulses with the OAM state can be generated by controlling the gain and cavity loss of each transverse mode. Therefore, generating OAM beams via harmonic amplification in a helical undulator is possible with an XFELO. However, the setup for harmonic amplification is complex in multi-pass facilities, and no practical helical undulator has been realized for short-wavelength free-electron lasers. Meanwhile, other methods for generating OAM beams focus on the suppression of the fundamental Gaussian mode by increasing the cavity loss. In such methods, instead of direct manipulation optical elements, an amplitude mask is used to realize OAM X-ray beams~\cite{Best1990,10.1063/1.345634,wu:fel2019}. However, the fabrication of high-precision amplitude masks may be difficult. Additionally, this approach is operation-wavelength-dependent as the diffraction effect varies with the X-ray wavelength. 

\begin{figure}
    \centering
    \subfigure{\includegraphics*[width=160pt]{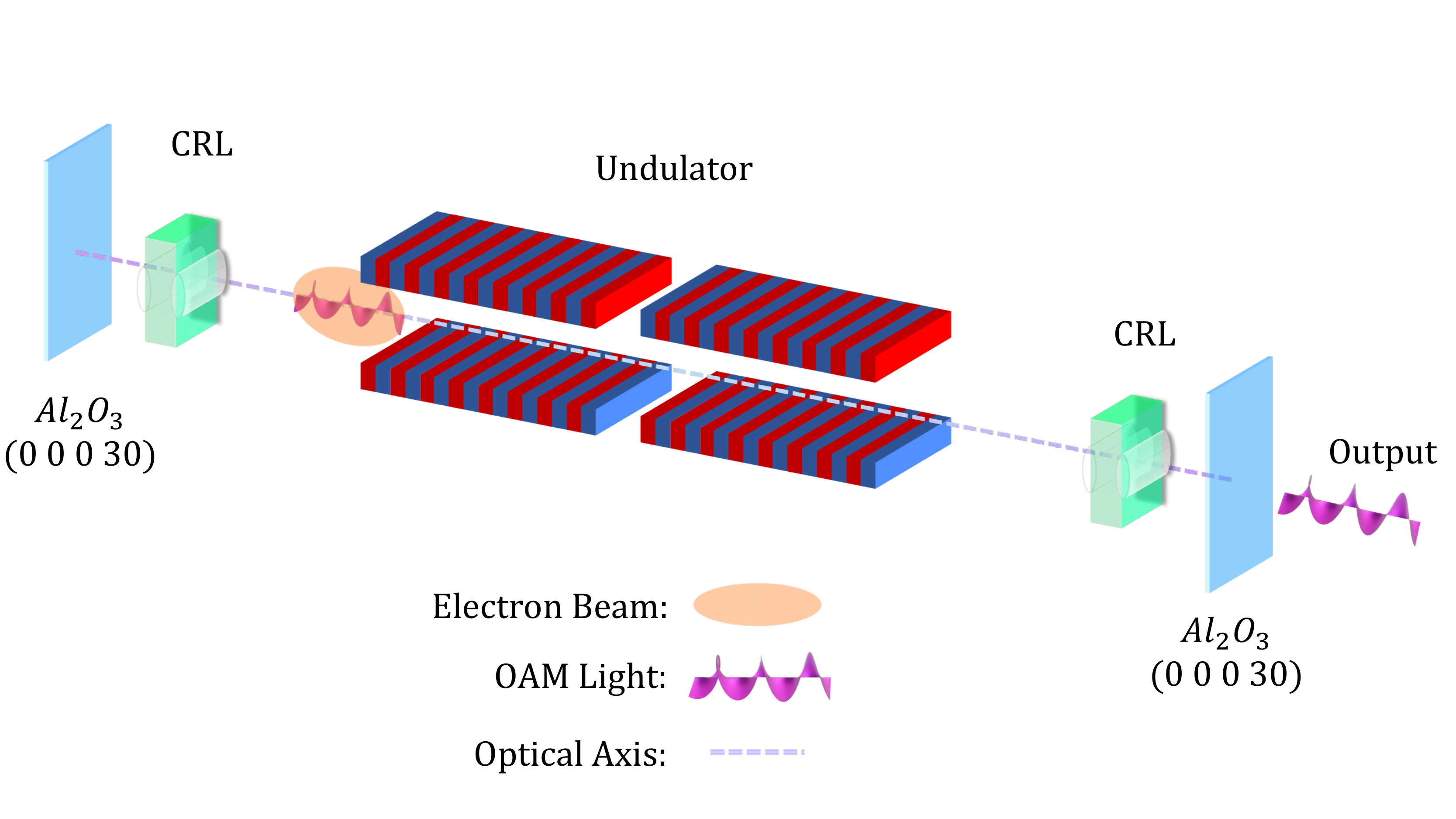}}
    \caption{Scheme to generate X-ray orbital angular momentum (OAM) beams by using X-ray free-electron laser oscillator.}
    \label{fig:scheme}
\end{figure}

Against this backdrop, herein, we present a novel and highly flexible method that essentially facilitates the preservation of OAM beam amplification at the fundamental wavelength and avoids the need for external optical elements. The schematic of our method is shown in Fig.~\ref{fig:scheme}. The method is very simple as it requires only the adjustment of the resonant condition of XFELO operation. The key element is the Bragg mirrors, which reflect X-rays via Bragg reflection with nearly 99\% reflectivity within a relative spectrum width of approximately $1\times10^{-6}$~\cite{Shvydko2010}. The method is based on the important fact that the resonant condition for each transverse mode is slightly different. As a result, the gain profile of each transverse mode is shifted over the spectrum. The combined effect is that XFELOs can operate in a specific spectral regime wherein the radiation in a high-order transverse mode can be obtained at maximum gain. Because the laser saturation state is only governed by the gain and cavity loss, this effect enables XFELOs to select transverse modes that carry OAM. This approach significantly reduces the number of electron-beam-control elements and the total number of external optical elements, thereby significant increasing the optical efficiency.



To derive a mode capable of conforming with the XFELO OAM selection rules, we start with the Laguerre--Gaussian (LG) modes that can possess arbitrary OAM~\cite{Allen1992}. With the LG basis modes, the transversely dominated FEL radiation fields can be described as $E(r, \phi, z) = \sum a_{p,l} u_{p,l}(r, \phi, z) \exp(ikz) $, where $u_{p,l}$ can be expressed as
\begin{equation}
     \begin{aligned}
     u_{p,l}(r, \phi, z)=& \frac{C_{p,l}}{w(z)}\left(\frac{r \sqrt{2}}{w(z)}\right)^{|l|} L_{p}^{|l|}\left[\frac{2 r^{2}}{w^{2}(z)}\right] \\
     & \times \exp \left[\frac{-r^{2}}{w^{2}(z)}\right] \exp \left[ \frac{-i k r^{2} z}{2\left(z^{2}+z_{R}^{2}\right)} \right] \exp  (-i l \phi) \\
     & \times \exp \left[i(2 p+l+1) \tan ^{-1} \frac{z}{z_{R}}\right]
     \end{aligned}
     \label{eq:LG}
\end{equation}
in which $C_{p,l} = \sqrt{2 p !/\pi(p+|l|) !}$ denotes a required normalization constant and $L_{p}^{l}(x)=\sum_{j=0}^{p}(p+l) !(-x)^{j} / j !(p-j) !(l+j) !$ is an associated Laguerre polynomial. Parameters $p$ and $l$ describe an array of unique modes, where $l$ denotes any real integer for the azimuthal mode and $p$ is zero or any positive integer for the radial mode. Additionally, $z_R = \pi w_0^2 / \lambda$ denotes the Rayleigh length with wavelength $\lambda$ and beam waist $w_0$, and the spot size parameter along the beam propagation is given by $w(z) = w_0(z^2/z_R^2 + 1)^{1/2}$. These optical parameters are governed by the cavity length and focusing-element configuration. Without special requirements, $w_0$ is designed to be located at the undulator center. 


In a low-gain device, the single-pass gain related to FEL amplification in the undulator is proportional to the overlap area between the electron beam and the radiation field. Thus, the so-called "filling factor" can be used to scale the gain value and, thus, to avoid solving the full equation for each expansion mode. The procedure outlined in Ref.~\cite{10.1103/physreva.77.063831} can be applied to the high-order LG modes. With a dimensionless electron beam density profile $f(r, \phi)$, the spatial mode coupling coefficient for any LG mode can be expressed as 
\begin{equation}
     \mathrm{F}_{l,p; l', p'}= \frac{ \iint f(r, \phi) u_{p', l'}(r, \phi) u*_{p, l}(r, \phi) r d \phi d r}  {\iint\left|u_{p, l}(r, \phi)\right|^{2} r d \phi d r}.
\end{equation}

\begin{figure}[!htb]
     \centering
     \subfigure{\includegraphics*[width=240pt]{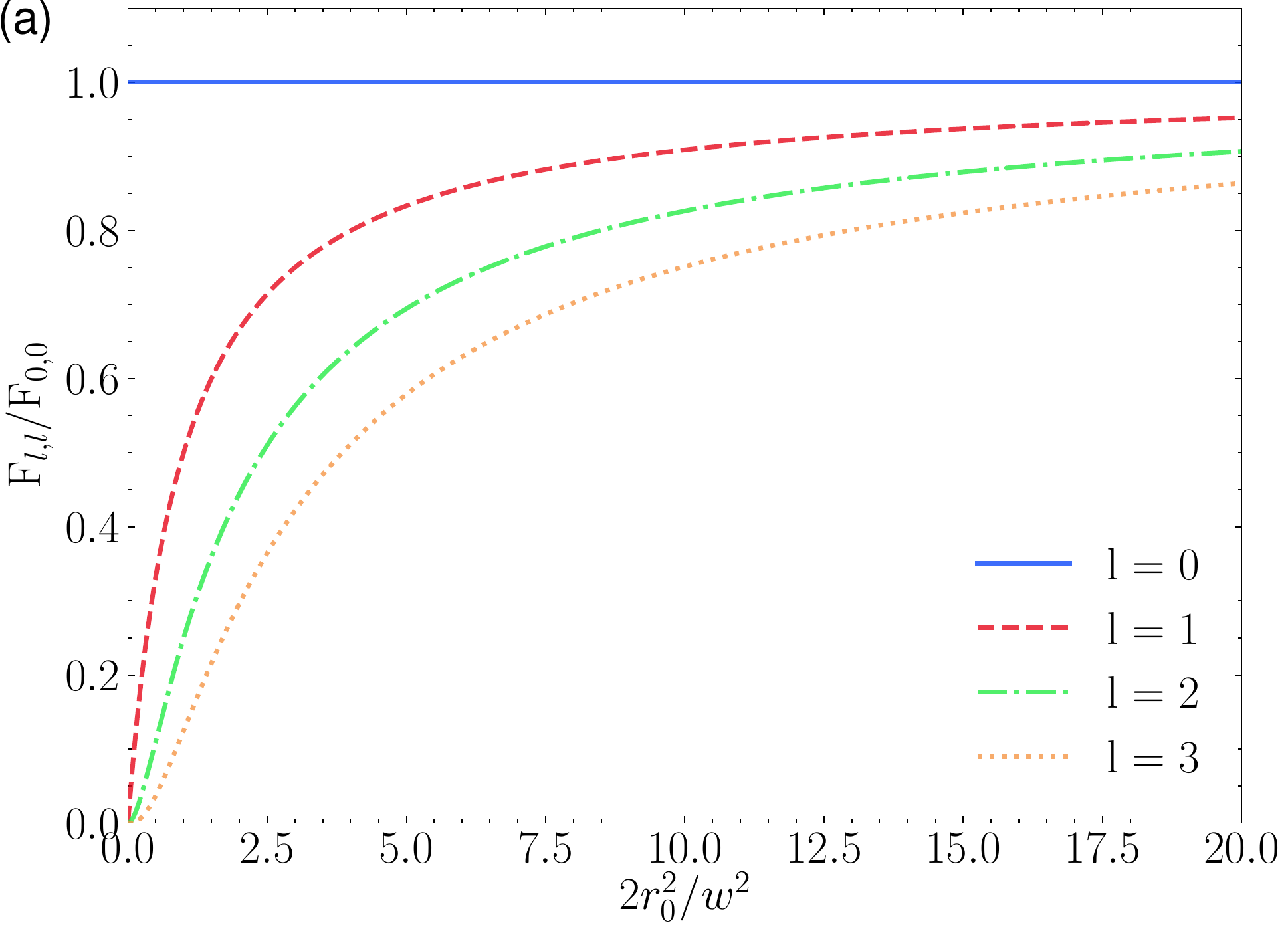}}
     \subfigure{\includegraphics*[width=240pt]{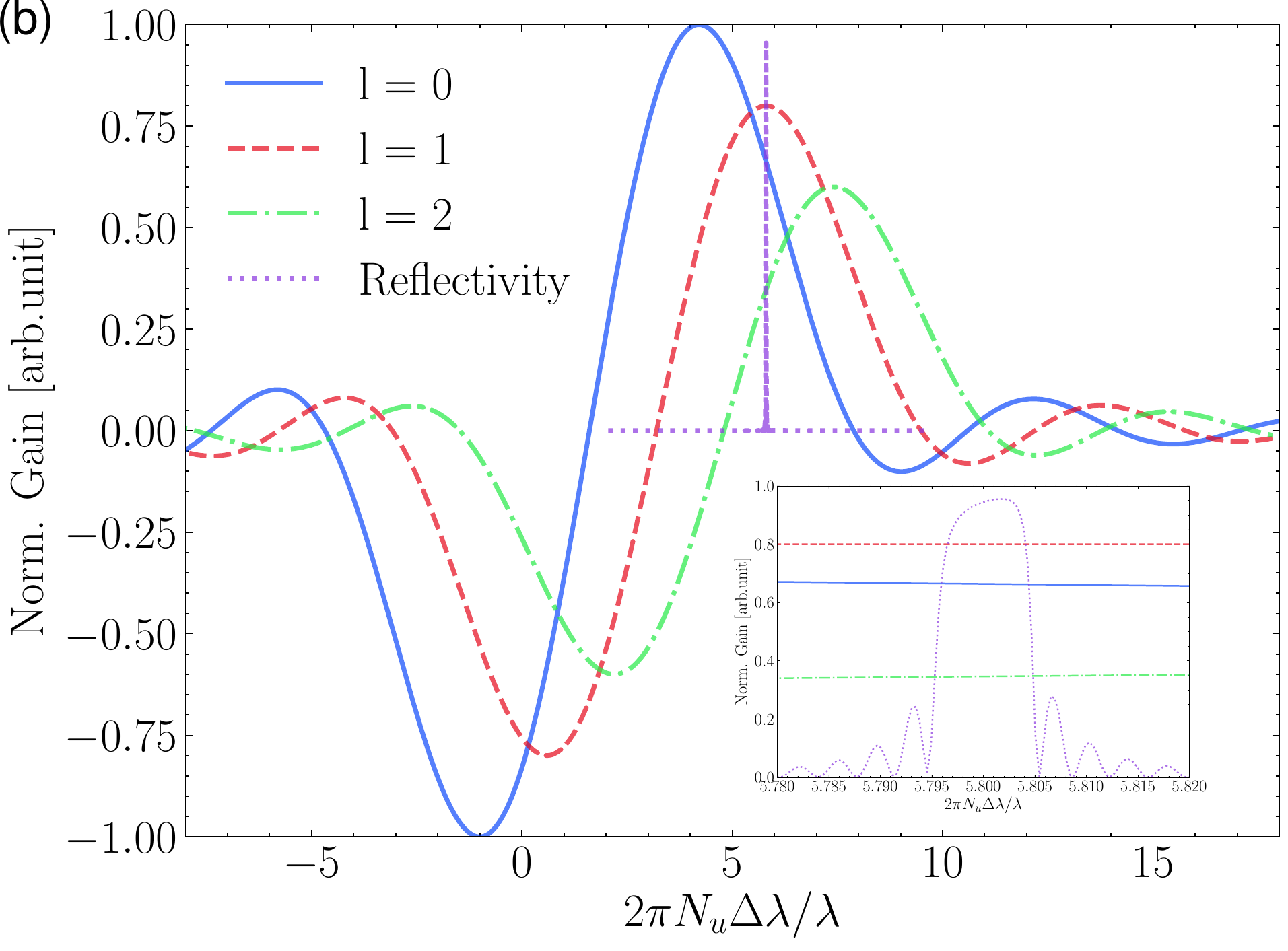}}
     \caption{Plot~(a) shows the variation in the normalized filling factor $\mathrm{F}_{l,l}$ for several azimuthal modes $l$. Plot~(b) shows the single-pass gain for different azimuthal modes $l$. The maximum gain of the Gaussian mode is normalized to one. For individual LG modes, the gain is scaled by the ratio of the filling factor and shifted by an amount equal to the difference in the resonant wavelength. The purple line in plot~(b) indicates the reflectivity of the Bragg mirror.}
     \label{fig:Ff}
\end{figure}

We consider only the higher azimuthal modes with the lowest-order radial mode $p=0$; otherwise, modes of $p>1$ would cascade down toward the fundamental radial mode with $p=0$ owing to the non-zero coupling efficiency between each radial mode~\cite{10.1103/physreva.77.063831}. Thus, the coupling coefficient can be simplified as $\mathrm{F}_{l;l'}$ without the radial mode, and $\mathrm{F}_{l;l}$ then denotes the commonly used filling factor~\cite{10.1007/bf00690020,10.1088/0963-9659/2/3/006}. In a realistic scenario, the Gaussian electron beam transverse density profile can be expressed as $f(r, \phi) = \exp(-r^2/r_0^2)$, where $r_0$ denotes the transverse radius. Consequently, $\mathrm{F}_{l;l'}$ can be integrated via Gaussian integrals as 
\begin{equation}
     \mathrm{F}_{l;l'} = \delta_{ll'} \left( \frac{x^2}{ 1+x^2 } \right)^{|l|+1}
     \label{eq:F_ll}
\end{equation}
where  $x = \sqrt{2}r_0/w$ is defined as the normalized electron beam radius. Parameter $\delta_{ll'}$ indicates that all the azimuthal modes are not coupled together, with an axisymmetric electron beam transverse profile. For convenience, the filling factor, $\mathrm{F}_{l;l}$, is normalized by $\mathrm{F}_{0;0}$. Fig.~\ref{fig:Ff}~(a) illustrates the normalized filling factor as a function of the electron beam size and the mode order $l$. 

The intensities of the modes with $|l|>0$ vanish on-axis, and the trend increases with increase in $l$. Therefore, the coupling factor between the electron beams and radiation decreases with $l$, as can be observed in Fig.~\ref{fig:Ff}~(a). For any radial size, the Gaussian mode corresponding to $l=0$ is found to yield the largest coupling factor when compared with those of the other azimuthal modes. This explains why X-ray pulses generated in the XFELO are limited to the Gaussian mode. In the case of larger electron beam sizes,  $\mathrm{F}_{l;l}$ increases and approaches unity, which means that radiation with $|l|>0$ can be amplified more equally in the interaction with the electron beam. Hence, an electron beam with a large radial size is preferred for the amplification of the high-order OAM mode.

However, there is a trade-off between the coupling efficiency and the single-pass gain. The single-pass gain can be approximated as
\begin{equation}
     G_{l} \approx \frac{4 \pi^{2} e^{2} N_u^3 \lambda_u^{2} K^{2}[J J]^{2}}{\gamma^{3} m c^{2}} n_{e} \mathrm{F}_{l,l} \cdot\left[\frac{2-2 \cos \nu_{}-\nu_{} \sin \nu_{}}{\nu_{}^{3}}\right]
     \label{eq:gain}
\end{equation}
where 
\begin{eqnarray}
     && [J J] \equiv J_{0}\left(\frac{K^{2}}{4+2 K^{2}}\right)-J_{1}\left(\frac{K^{2}}{4+2 K^{2}}\right), \\
     && n_e = I/ e c A_e, \quad A_e = \iint f(r, \phi) dr d\phi,
\end{eqnarray}
where $A_e$ denotes the electron beam area~\cite{10.1007/bf00690020}. In the above equations, $N_u$ denotes the number of undulator periods, $K$ is the undulator parameter, $\lambda_u$ is the length of the undulator periods, $I$ is the current, $c$ is the speed of light, $\gamma$ is the electron energy, $m$ is the rest mass of the electron, and $\nu$ is the resonant deviation. From Eq.~\ref{eq:gain}, we can infer that a large single-pass gain corresponds to a large electron density $n_e$ resulting from a small $x$ value. In practice, $x^2 = |l|+1$ is a good approximation when one of the high-order azimuthal modes is selected for amplification~\cite{Padgett2015}.

     

In Eq.~\ref{eq:LG}, the extra phase term $(2p +|l|+1) \tan ^{-1} z/z_{R}$, which refers to the Gouy phase, is associated with mode order $l$. The presence of the Gouy phase results in a difference in the phase velocity. Thus, the phase of the ponderomotive potential in the FEL equations is associated with the individual LG modes, which means that the FEL resonant wavelength is slightly different for each high-order LG mode. The averaged deviation~\cite{10.1088/0963-9659/2/3/006,Classical2014} is 
\begin{equation}
\left\langle\frac{\Delta \lambda}{\lambda}\right\rangle \approx \frac{2(2p+|l|+1) \tan ^{-1}\left(L_{u} / 2 z_R \right)}{2 \pi N_{u}},
\end{equation}
where $p$ is zero in our consideration, and $L_u$ denotes the undulator length. In the case of resonant deviation, $\nu$ is given by $\nu = \nu_0 + 2(|l|+1) \tan ^{-1}\left(L_{u} / 2 z_R \right)$. In this case, the shape of the gain profile is identical, while the gain profile of the individual LG modes is scaled by the filling factor and is shifted by an amount equal to the difference in the resonant wavelength, as shown in Fig.~\ref{fig:Ff}~(b).

\begin{figure}[!htb]
     \centering
     \subfigure{\includegraphics*[width=240pt]{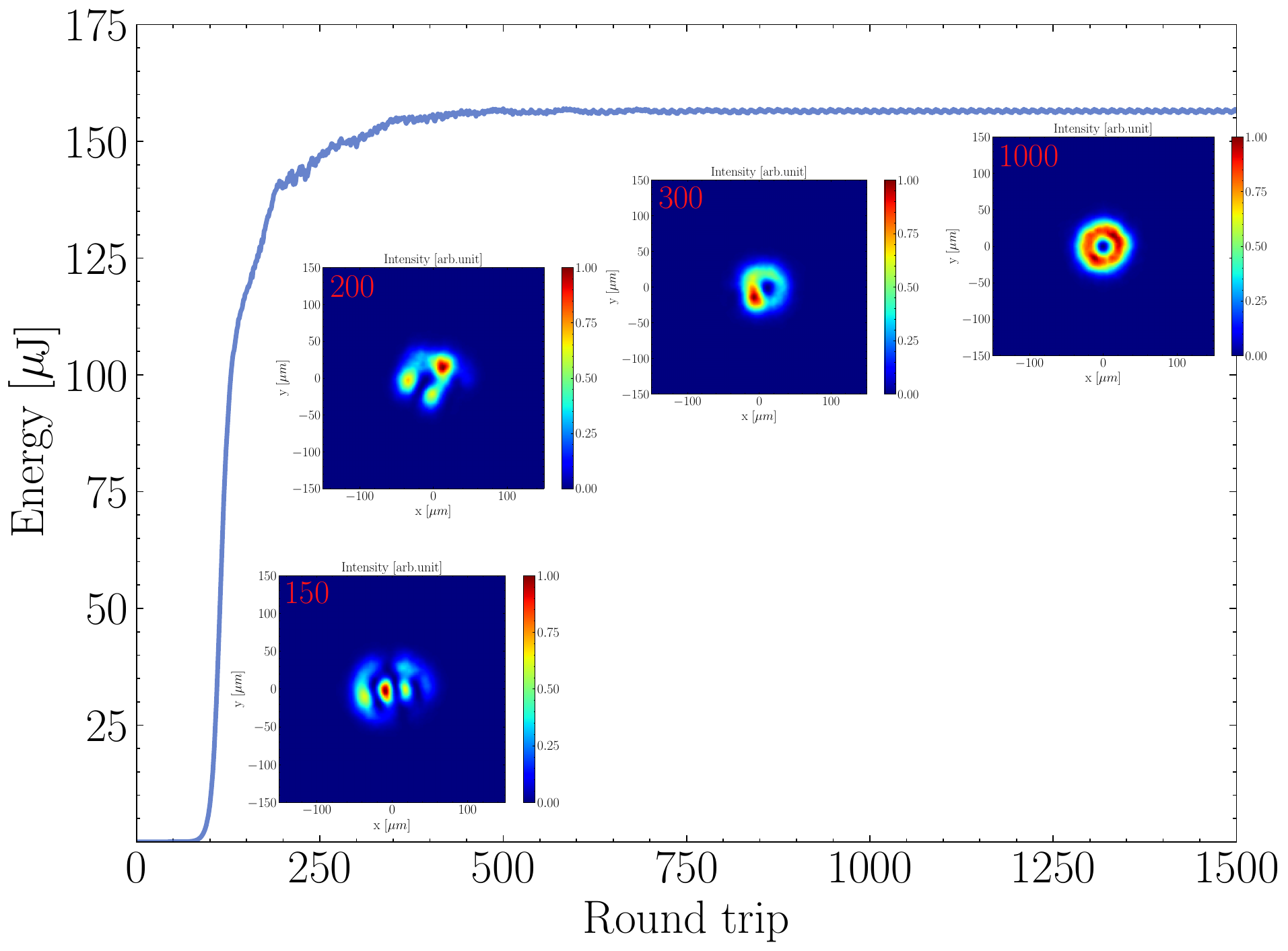}}
     \caption{Power growth for $l=1$ mode. The transverse intensity profiles for the round trips of 150, 200, 300, and 1000 are shown. The evolution of the intensity profile into a doughnut-like one can be clearly observed.}
     \label{fig:power}
\end{figure}
A characteristic of the XFELO that its relative spectral bandwidth is only $1\times10^{-6}$. This is because the XFELO mirrors act as high-resolution filters to identical degrees across the spectrum. Thus, the use of a crystal mirror can confer the ability to precisely select the XFELO working wavelength regime. As shown in Fig.~\ref{fig:Ff}, the appropriate selection of the wavelength regime can maximize the gain of the LG mode with $l=1$, thereby amplifying the OAM beams in the XFELO with stability and purity. In other words, the Bragg crystal can act as both a mirror and "mode selector". Consequently, it is possible to select the mode by simply controlling the resonant wavelength deviation.

We performed numerical simulations of OAM beam generation in an XFELO with the proposed scheme based on the parameters applied in the Shanghai High-Repetition-Rate XFEL and Extreme Light Facility (SHINE), which is the first hard X-ray FEL facility in China (currently under construction)~\cite{Yan2019,Li2018SCLF}. Here, we note that a continuous wave (CW) superconducting linac at SHINE can deliver 8~GeV electron bunches with ultra-low 0.4~$\mu$m~rad normalized emittance at a 1~MHz repetition rate; these settings are suitable for XFELO operation. Additionally, the electron bunch with a total charge of 100~pC is compressed to a peak current of 0.5~kA. The resonant X-ray photon energy is set to approximately 14.3~keV, which is equal to the Bragg energy of sapphire crystal mirrors at normal incidence to the (0 0 0 30) atomic planes. A typical FODO lattice is employed to control the electron beam size. In the study, three-dimensional simulations were carried out by using a combination of GENESIS~\cite{Reiche1999}, OPC~\cite{VanDerSlot2009} and BRIGHT~\cite{Huang2019}. In addition, four undulator segments were used to provide sufficient gain. In the simulation, the resonant wavelength deviation $\Delta \lambda/ \lambda$ was 0.11\%, which is preferable for the OAM mode with $l=1$. 


\begin{figure}
     \centering
     \subfigure{\includegraphics*[width=120pt]{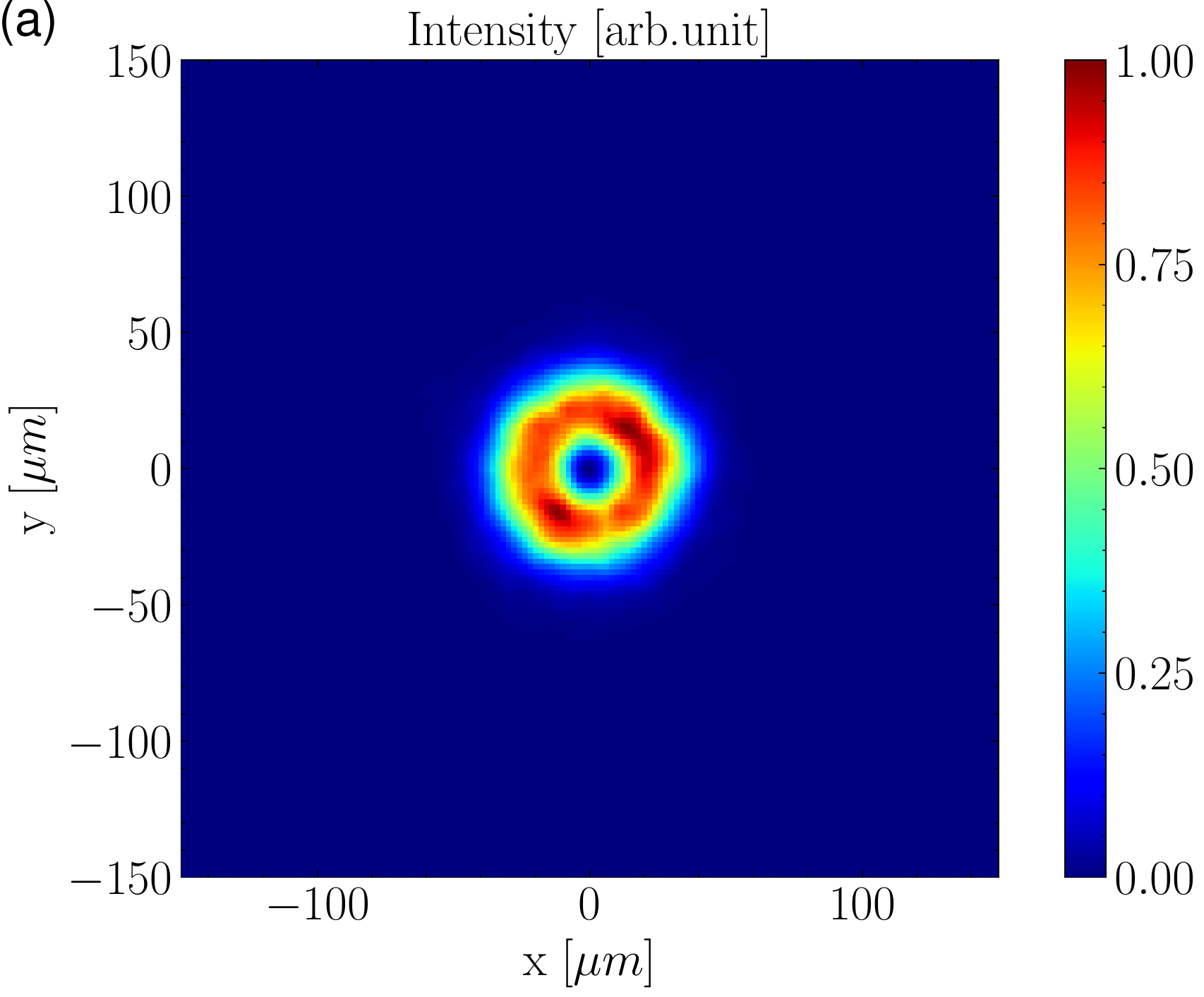}}
     \subfigure{\includegraphics*[width=120pt]{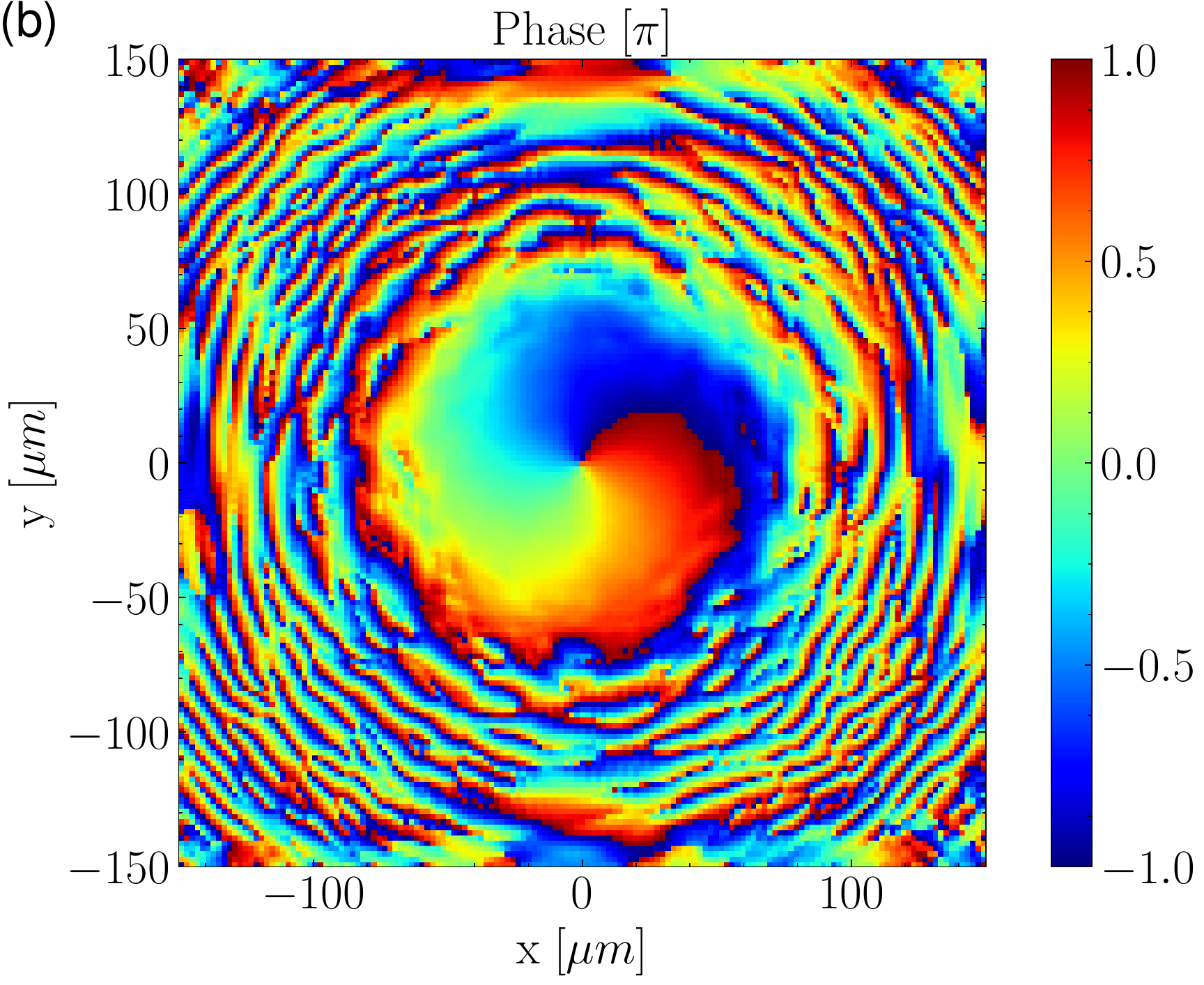}}
     \subfigure{\includegraphics*[width=120pt]{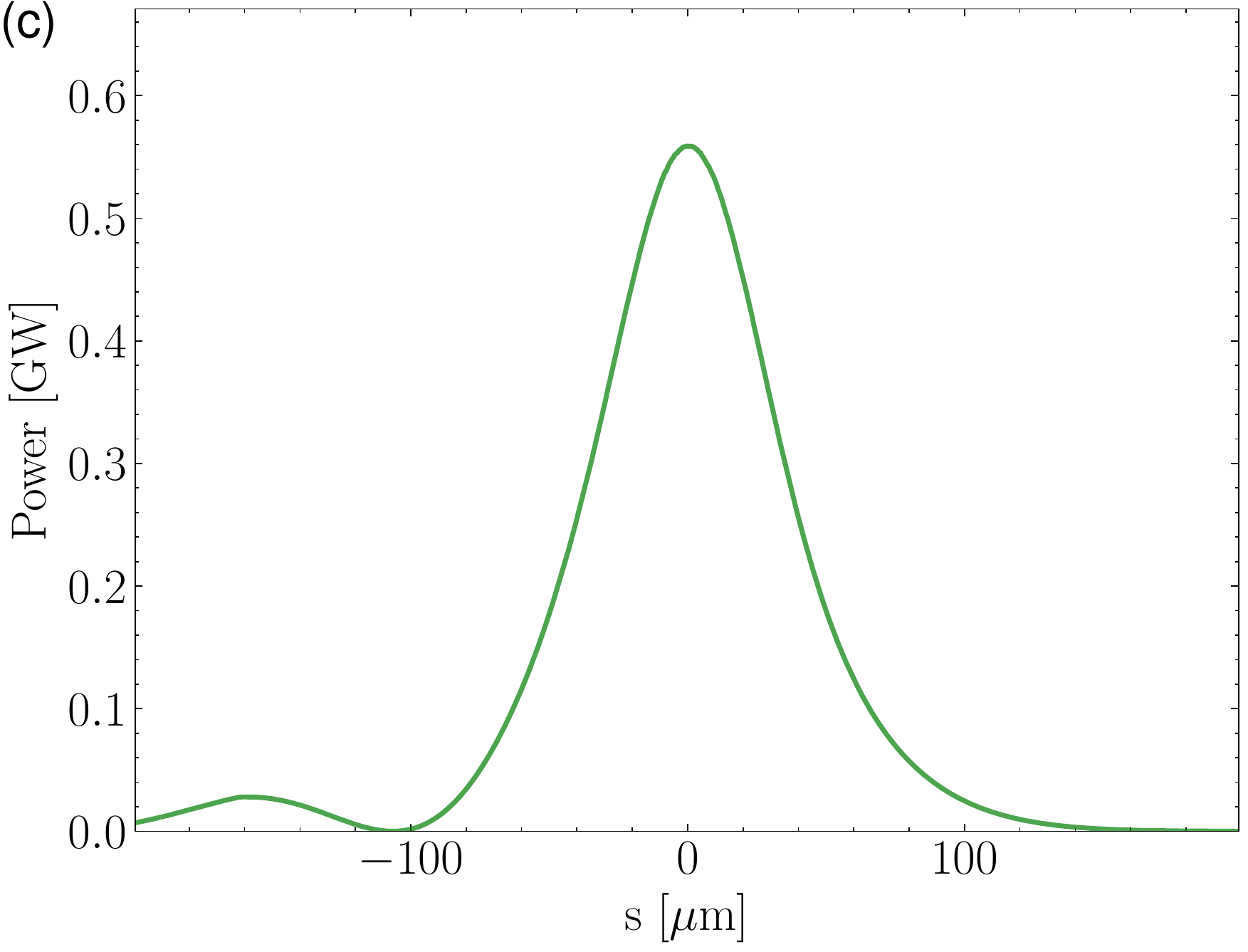}}
     \subfigure{\includegraphics*[width=120pt]{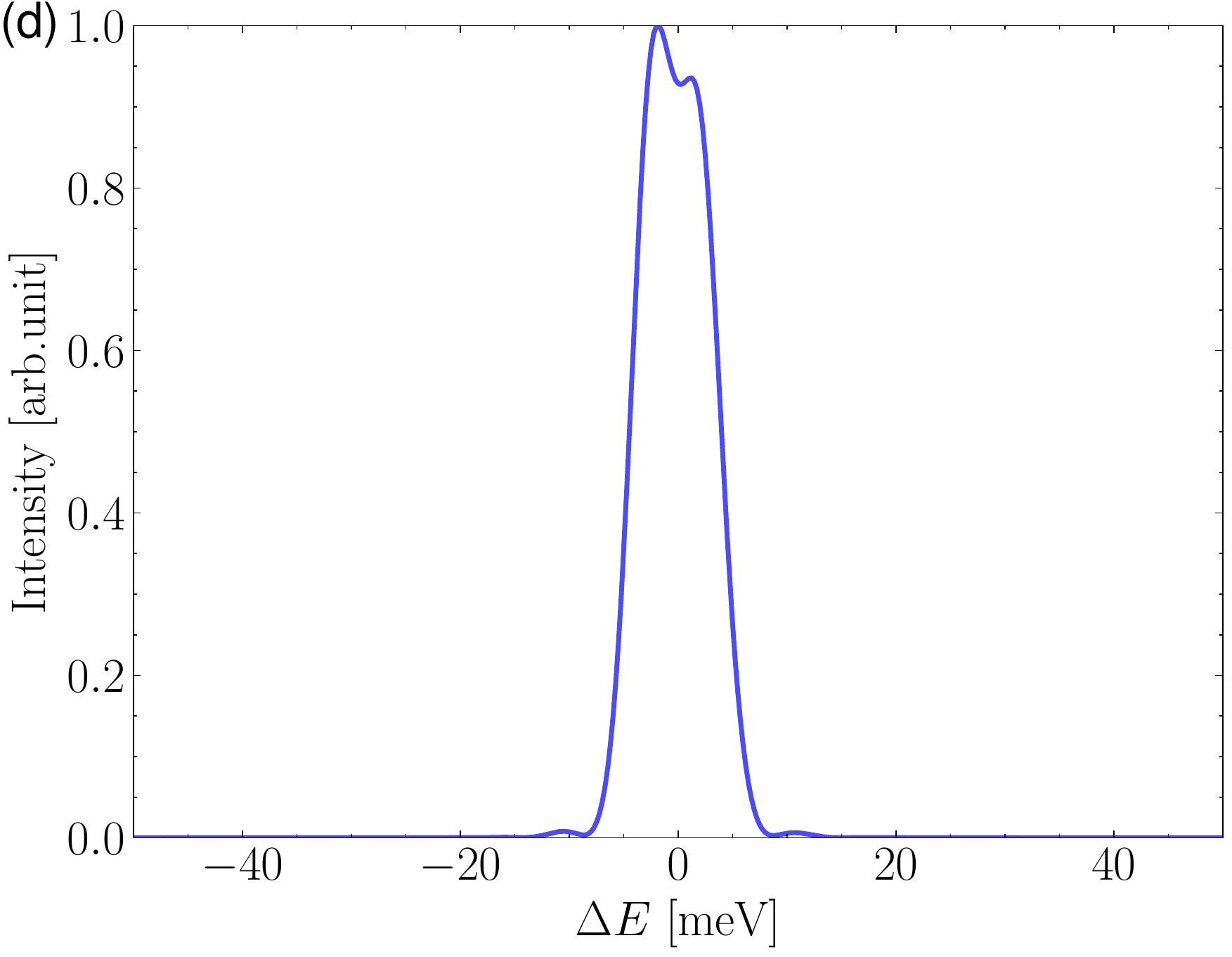}}
     \caption{Transverse intensities (a), corresponding phase distribution (b), longitudinal power profile (c) and its corresponding spectrum (d). The transverse profile of the light reveals a dominant OAM mode of $l=1$ at saturation. A peak power of 500 MW with a full-width at half-maximum (FWHM) spectral width of 11 meV can be observed.}
     \label{fig:trans}
\end{figure}

The simulation results are shown in Fig.~\ref{fig:power} and Fig.~\ref{fig:trans}. Fig.~\ref{fig:power} shows the pulse energy as a function of the number of round trips and the evolution of the transverse mode. The saturation power reaches 150~$\mu$J, approaching the level of Gaussian mode operation. In the short period ranging from 150 to 400 round trips, mode competition or mode completion is observed to occur because the transverse mode during exponential gain is an asymmetric mixture, as can be observed in the subplot~(150) of Fig.~\ref{fig:power}. This mixture is unstable as it cannot reach the maxima of the filling factor. As a consequence, a short time period is required for completing the symmetric doughnut-like transverse intensity profile and the helical phase. The outcome is robust because the system must reach a stable state determined by the gain that is tuned for the mode with $l=1$.




At saturation, the characteristic hollow profile and helical phase can be observed in Fig.~\ref{fig:trans}. It is obvious from the phase distribution that the dominant mode is the $l=1, p=0$ LG mode. The intensity of modes $l \neq 1$ is negligible. The longitudinal power profile and the spectrum are shown in Fig.~\ref{fig:trans}. While the total power increases up to 150~$\mu$J, the peak power exceeds 500~MW with the FWHM spectrum width of 11 meV. These results regarding the generation of the $l=1$ OAM mode at the fundamental frequency of the XFELO suggest that resonant deviation in an XFELO can be partly used to control the transverse mode.

\begin{figure}
    \centering
    \subfigure{\includegraphics*[width=120pt]{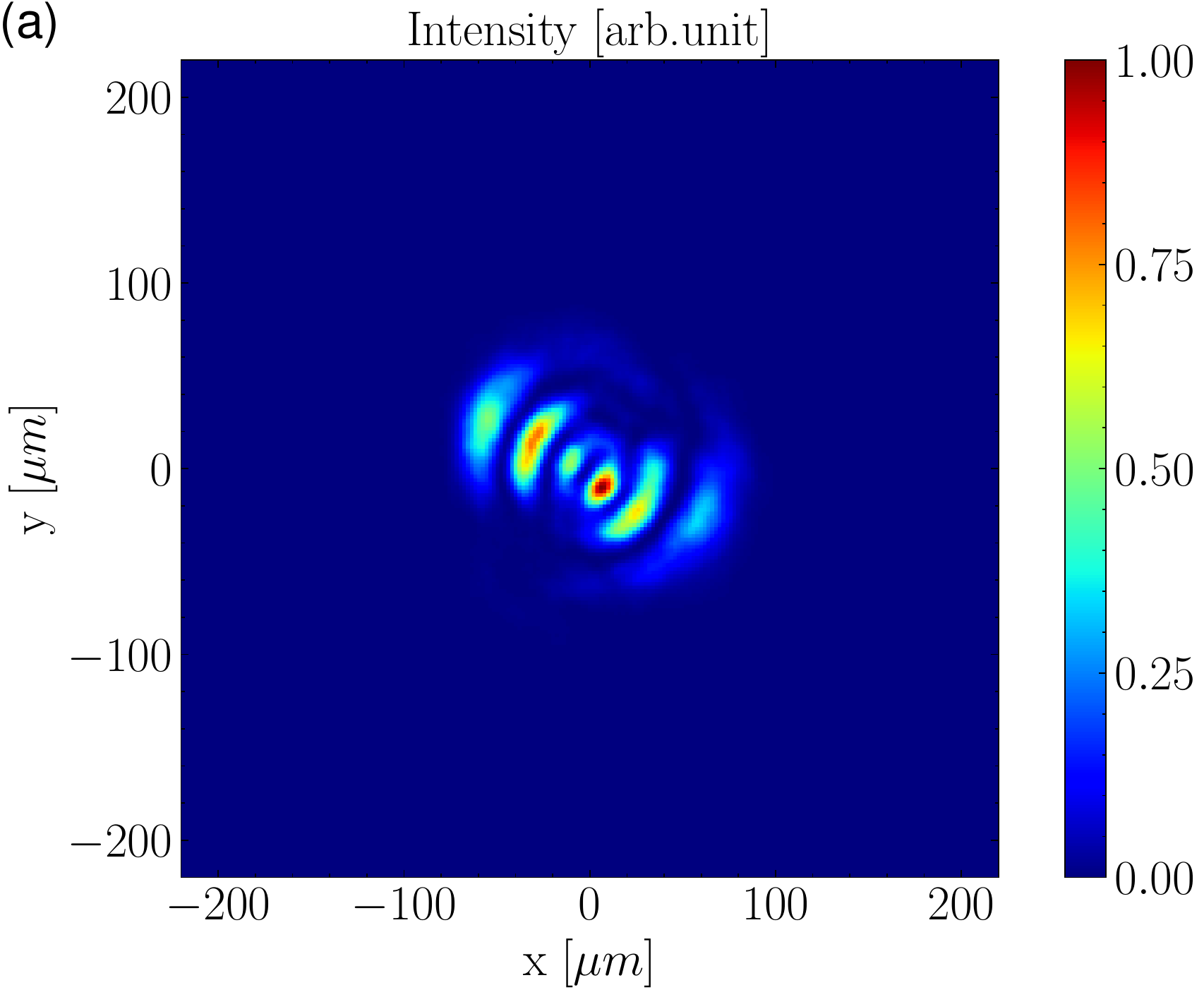}}
    \subfigure{\includegraphics*[width=120pt]{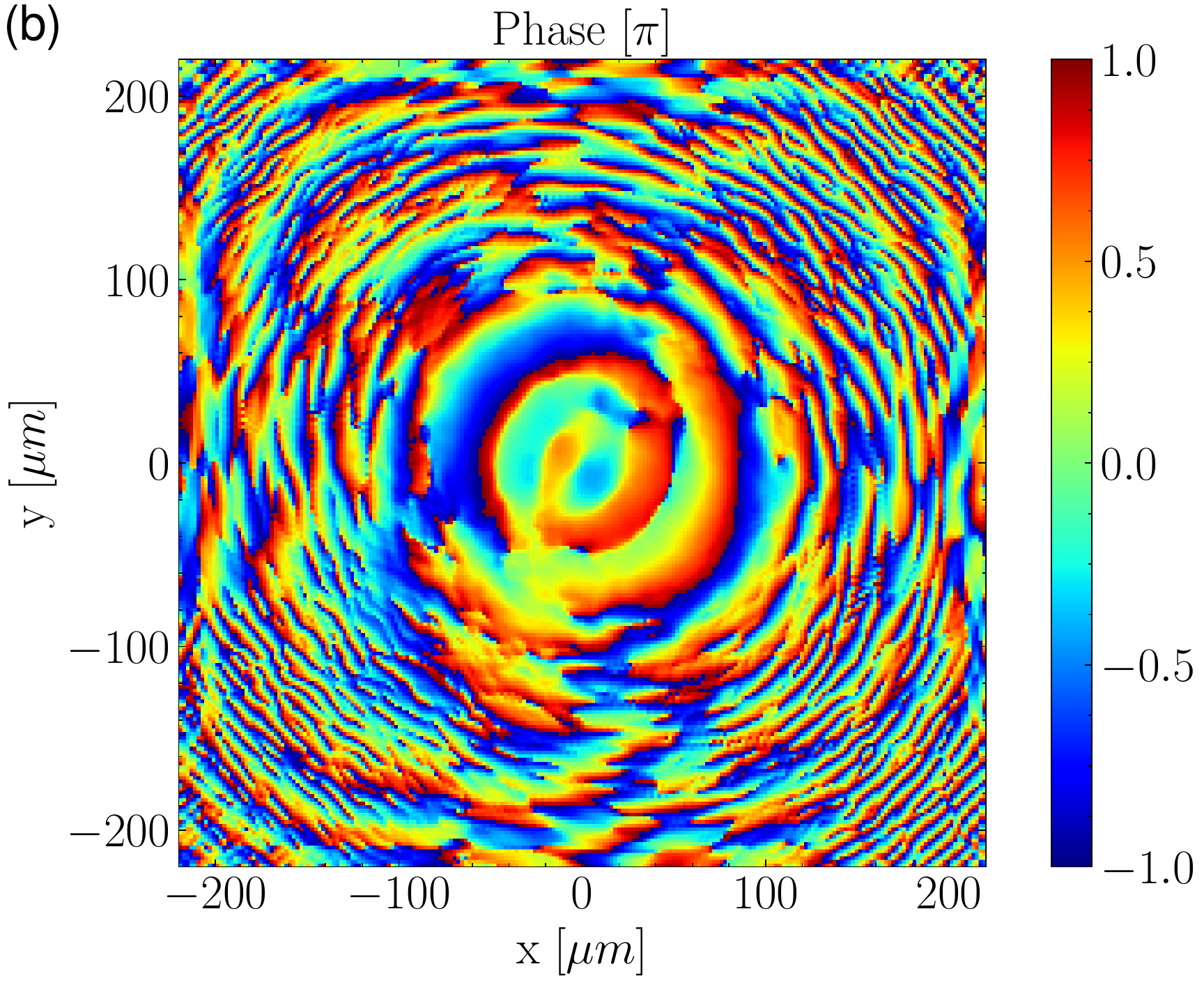}}
    \caption{Transverse profile (a) and transverse phase distribution (b) when the resonant deviation is selected to amplify the $l=2$ mode. The corresponding $\Delta \lambda/ \lambda$ is 0.143\%}
    \label{fig:transl2}
\end{figure}

However, the most difficult challenge is the generation of higher-order transverse modes with purity in saturation. The key to select the OAM mode is the deviation in the resonant condition for each transverse mode owing to the Gouy phase. In this case, the HG modes can also be amplified, because they provide a similar Gouy phase, which is given by $(m+n+1) \tan ^{-1} z/z_{R}$ where integers $m,n$ indicate the mode ~\cite{Allen1992}. When $l$ is determined, a specific set of HG modes is amplified, for e.g., $l=2$ for HG$_{02}$, HG$_{20}$, and HG$_{11}$. Thus, a mixed mode may be observed because this can yield a larger filling factor. This effect is trivial in the mode with $|l|=1$, for its filling factor is larger than those of all the mixed high-order modes. Nevertheless, this effect can be a problem for the $|l|>1$ modes. This is largely because the zero central intensities increase with $|l|$, and the mixture of HG and LG modes can yield the maximum coupling efficiency leading to maximum gain. As a result, when a resonant deviation is selected to amplify the $|l|>1$ modes, the system will evolve to a state characterized by a complex transverse mode mixed with higher HG and LG modes, as can be observed in Fig.~\ref{fig:transl2}. Thus, the mixed mode at saturation is difficult to predict. Therefore, the approach to increase the mode purity is to use a crystal mirror with a hole. The hole can increase the cavity loss of unwanted modes and subsequently suppress the amplification of these modes. Additionally, the hole can be replaced by an unpolished area over the surface or a defect in the crystal.

In conclusion, we proposed the OAM operation of an XFELO without the need for external optical elements. We applied a theoretical analysis based on the mode--electron coupling factor and Gouy phase shift to describe the gain profile evolution. In the proposed method, the Bragg crystal serves not only as a reflector but also as a "mode selector" both in the longitudinal and transverse modes, which enables the specified resonance deviation that maximizes the single-pass gain of the high-order OAM mode of interest. This allows the amplification and saturation of OAM beams in the typical XFELO configuration. The feasibility of the proposed design was verified through three-dimensional numerical simulations based on the SHINE parameters. The typical results for the two-mirror-symmetry cavity indicated that the OAM operation of an XFELO can generate 150~$\mu$J pulses in the $l=1$ mode. Without the need for any additional elements, the proposed method demonstrates the simplest approach to generate the $|l|=1$ OAM mode in an XFELO. The limitation of this method is the difficulty to generate the $|l|>1$ OAM modes without optical spatial filtering. We believe that our approach can promote future applications of X-ray free-electron lasers.

\begin{acknowledgments}
This work was partially supported by the National Key Research and Development Program of China (Grant Numbers 2018YFE0103100, 2016YFA0401900) and the National Natural Science Foundation of China (Grant Numbers 11935020, 11775293).
\end{acknowledgments}


\bibliographystyle{apsrev4-1}


\begin{thebibliography}{37}%
     \makeatletter
     \providecommand \@ifxundefined [1]{%
      \@ifx{#1\undefined}
     }%
     \providecommand \@ifnum [1]{%
      \ifnum #1\expandafter \@firstoftwo
      \else \expandafter \@secondoftwo
      \fi
     }%
     \providecommand \@ifx [1]{%
      \ifx #1\expandafter \@firstoftwo
      \else \expandafter \@secondoftwo
      \fi
     }%
     \providecommand \natexlab [1]{#1}%
     \providecommand \enquote  [1]{``#1''}%
     \providecommand \bibnamefont  [1]{#1}%
     \providecommand \bibfnamefont [1]{#1}%
     \providecommand \citenamefont [1]{#1}%
     \providecommand \href@noop [0]{\@secondoftwo}%
     \providecommand \href [0]{\begingroup \@sanitize@url \@href}%
     \providecommand \@href[1]{\@@startlink{#1}\@@href}%
     \providecommand \@@href[1]{\endgroup#1\@@endlink}%
     \providecommand \@sanitize@url [0]{\catcode `\\12\catcode `\$12\catcode
       `\&12\catcode `\#12\catcode `\^12\catcode `\_12\catcode `\%12\relax}%
     \providecommand \@@startlink[1]{}%
     \providecommand \@@endlink[0]{}%
     \providecommand \url  [0]{\begingroup\@sanitize@url \@url }%
     \providecommand \@url [1]{\endgroup\@href {#1}{\urlprefix }}%
     \providecommand \urlprefix  [0]{URL }%
     \providecommand \Eprint [0]{\href }%
     \providecommand \doibase [0]{http://dx.doi.org/}%
     \providecommand \selectlanguage [0]{\@gobble}%
     \providecommand \bibinfo  [0]{\@secondoftwo}%
     \providecommand \bibfield  [0]{\@secondoftwo}%
     \providecommand \translation [1]{[#1]}%
     \providecommand \BibitemOpen [0]{}%
     \providecommand \bibitemStop [0]{}%
     \providecommand \bibitemNoStop [0]{.\EOS\space}%
     \providecommand \EOS [0]{\spacefactor3000\relax}%
     \providecommand \BibitemShut  [1]{\csname bibitem#1\endcsname}%
     \let\auto@bib@innerbib\@empty
     \bibitem [{\citenamefont {Lee}\ \emph {et~al.}(2019)\citenamefont {Lee},
       \citenamefont {Alexander}, \citenamefont {Kevan}, \citenamefont {Roy},\ and\
       \citenamefont {McMorran}}]{10.1038/s41566-018-0328-8}%
       \BibitemOpen
       \bibfield  {author} {\bibinfo {author} {\bibfnamefont {J.~C.}\ \bibnamefont
       {Lee}}, \bibinfo {author} {\bibfnamefont {S.~J.}\ \bibnamefont {Alexander}},
       \bibinfo {author} {\bibfnamefont {S.~D.}\ \bibnamefont {Kevan}}, \bibinfo
       {author} {\bibfnamefont {S.}~\bibnamefont {Roy}}, \ and\ \bibinfo {author}
       {\bibfnamefont {B.~J.}\ \bibnamefont {McMorran}},\ }\href {\doibase
       10.1038/s41566-018-0328-8} {\bibfield  {journal} {\bibinfo  {journal} {Nat.
       Photonics}\ }\textbf {\bibinfo {volume} {13}},\ \bibinfo {pages} {205}
       (\bibinfo {year} {2019})}\BibitemShut {NoStop}%
     \bibitem [{\citenamefont {Shen}\ \emph {et~al.}(2019)\citenamefont {Shen},
       \citenamefont {Wang}, \citenamefont {Xie}, \citenamefont {Min} \emph
       {et~al.}}]{Shen2019}%
       \BibitemOpen
       \bibfield  {author} {\bibinfo {author} {\bibfnamefont {Y.}~\bibnamefont
       {Shen}}, \bibinfo {author} {\bibfnamefont {X.}~\bibnamefont {Wang}}, \bibinfo
       {author} {\bibfnamefont {Z.}~\bibnamefont {Xie}}, \bibinfo {author}
       {\bibfnamefont {C.}~\bibnamefont {Min}},  \emph {et~al.},\ }\href {\doibase
       10.1038/s41377-019-0194-2} {\bibfield  {journal} {\bibinfo  {journal} {Light
       Sci. Appl.}\ }\textbf {\bibinfo {volume} {8}} (\bibinfo {year} {2019}),\
       10.1038/s41377-019-0194-2}\BibitemShut {NoStop}%
     \bibitem [{\citenamefont {Sroor}\ \emph {et~al.}(2020)\citenamefont {Sroor},
       \citenamefont {Huang}, \citenamefont {Sephton}, \citenamefont {Naidoo} \emph
       {et~al.}}]{Sroor2020}%
       \BibitemOpen
       \bibfield  {author} {\bibinfo {author} {\bibfnamefont {H.}~\bibnamefont
       {Sroor}}, \bibinfo {author} {\bibfnamefont {Y.~W.}\ \bibnamefont {Huang}},
       \bibinfo {author} {\bibfnamefont {B.}~\bibnamefont {Sephton}}, \bibinfo
       {author} {\bibfnamefont {D.}~\bibnamefont {Naidoo}},  \emph {et~al.},\ }\href
       {\doibase 10.1038/s41566-020-0623-z} {\bibfield  {journal} {\bibinfo
       {journal} {Nat. Photonics}\ } (\bibinfo {year} {2020}),\
       10.1038/s41566-020-0623-z}\BibitemShut {NoStop}%
     \bibitem [{\citenamefont {Allen}\ \emph {et~al.}(1992)\citenamefont {Allen},
       \citenamefont {Beijersbergen}, \citenamefont {Spreeuw},\ and\ \citenamefont
       {Woerdman}}]{Allen1992}%
       \BibitemOpen
       \bibfield  {author} {\bibinfo {author} {\bibfnamefont {L.}~\bibnamefont
       {Allen}}, \bibinfo {author} {\bibfnamefont {M.~W.}\ \bibnamefont
       {Beijersbergen}}, \bibinfo {author} {\bibfnamefont {R.~J.}\ \bibnamefont
       {Spreeuw}}, \ and\ \bibinfo {author} {\bibfnamefont {J.~P.}\ \bibnamefont
       {Woerdman}},\ }\href {\doibase 10.1103/PhysRevA.45.8185} {\bibfield
       {journal} {\bibinfo  {journal} {Phys. Rev. A}\ }\textbf {\bibinfo {volume}
       {45}},\ \bibinfo {pages} {8185} (\bibinfo {year} {1992})}\BibitemShut
       {NoStop}%
     \bibitem [{\citenamefont {Sakdinawat}\ and\ \citenamefont
       {Liu}(2007)}]{Sakdinawat2007}%
       \BibitemOpen
       \bibfield  {author} {\bibinfo {author} {\bibfnamefont {A.}~\bibnamefont
       {Sakdinawat}}\ and\ \bibinfo {author} {\bibfnamefont {Y.}~\bibnamefont
       {Liu}},\ }\href {\doibase 10.1364/ol.32.002635} {\bibfield  {journal}
       {\bibinfo  {journal} {Opt. Lett.}\ }\textbf {\bibinfo {volume} {32}},\
       \bibinfo {pages} {2635} (\bibinfo {year} {2007})}\BibitemShut {NoStop}%
     \bibitem [{\citenamefont {{Van Veenendaal}}\ and\ \citenamefont
       {McNulty}(2007)}]{VanVeenendaal2007}%
       \BibitemOpen
       \bibfield  {author} {\bibinfo {author} {\bibfnamefont {M.}~\bibnamefont {{Van
       Veenendaal}}}\ and\ \bibinfo {author} {\bibfnamefont {I.}~\bibnamefont
       {McNulty}},\ }\href {\doibase 10.1103/PhysRevLett.98.157401} {\bibfield
       {journal} {\bibinfo  {journal} {Phys. Rev. Lett.}\ }\textbf {\bibinfo
       {volume} {98}} (\bibinfo {year} {2007}),\
       10.1103/PhysRevLett.98.157401}\BibitemShut {NoStop}%
     \bibitem [{\citenamefont {Peele}\ \emph {et~al.}(2002)\citenamefont {Peele},
       \citenamefont {McMahon}, \citenamefont {Paterson} \emph
       {et~al.}}]{Peele2002}%
       \BibitemOpen
       \bibfield  {author} {\bibinfo {author} {\bibfnamefont {A.~G.}\ \bibnamefont
       {Peele}}, \bibinfo {author} {\bibfnamefont {P.~J.}\ \bibnamefont {McMahon}},
       \bibinfo {author} {\bibfnamefont {D.}~\bibnamefont {Paterson}},  \emph
       {et~al.},\ }\href {\doibase 10.1364/ol.27.001752} {\bibfield  {journal}
       {\bibinfo  {journal} {Opt. Lett.}\ }\textbf {\bibinfo {volume} {27}},\
       \bibinfo {pages} {1752} (\bibinfo {year} {2002})}\BibitemShut {NoStop}%
     \bibitem [{\citenamefont {Vila-Comamala}\ \emph {et~al.}(2014)\citenamefont
       {Vila-Comamala}, \citenamefont {Sakdinawat},\ and\ \citenamefont
       {Guizar-Sicairos}}]{Vila-Comamala2014}%
       \BibitemOpen
       \bibfield  {author} {\bibinfo {author} {\bibfnamefont {J.}~\bibnamefont
       {Vila-Comamala}}, \bibinfo {author} {\bibfnamefont {A.}~\bibnamefont
       {Sakdinawat}}, \ and\ \bibinfo {author} {\bibfnamefont {M.}~\bibnamefont
       {Guizar-Sicairos}},\ }\href {\doibase 10.1364/ol.39.005281} {\bibfield
       {journal} {\bibinfo  {journal} {Opt. Lett.}\ }\textbf {\bibinfo {volume}
       {39}},\ \bibinfo {pages} {5281} (\bibinfo {year} {2014})}\BibitemShut
       {NoStop}%
     \bibitem [{\citenamefont {Emma}\ \emph {et~al.}(2010)\citenamefont {Emma},
       \citenamefont {Akre}, \citenamefont {Arthur}, \citenamefont {Bionta} \emph
       {et~al.}}]{emma2010first}%
       \BibitemOpen
       \bibfield  {author} {\bibinfo {author} {\bibfnamefont {P.}~\bibnamefont
       {Emma}}, \bibinfo {author} {\bibfnamefont {R.}~\bibnamefont {Akre}}, \bibinfo
       {author} {\bibfnamefont {J.}~\bibnamefont {Arthur}}, \bibinfo {author}
       {\bibfnamefont {R.}~\bibnamefont {Bionta}},  \emph {et~al.},\ }\href
       {\doibase 10.1038/nphoton.2010.176} {\bibfield  {journal} {\bibinfo
       {journal} {Nat. Photonics}\ }\textbf {\bibinfo {volume} {4}},\ \bibinfo
       {pages} {641} (\bibinfo {year} {2010})}\BibitemShut {NoStop}%
     \bibitem [{\citenamefont {Ishikawa}\ \emph {et~al.}(2012)\citenamefont
       {Ishikawa}, \citenamefont {Aoyagi}, \citenamefont {Asaka}, \citenamefont
       {Asano} \emph {et~al.}}]{ishikawa2012compact}%
       \BibitemOpen
       \bibfield  {author} {\bibinfo {author} {\bibfnamefont {T.}~\bibnamefont
       {Ishikawa}}, \bibinfo {author} {\bibfnamefont {H.}~\bibnamefont {Aoyagi}},
       \bibinfo {author} {\bibfnamefont {T.}~\bibnamefont {Asaka}}, \bibinfo
       {author} {\bibfnamefont {Y.}~\bibnamefont {Asano}},  \emph {et~al.},\
       }\href@noop {} {\bibfield  {journal} {\bibinfo  {journal} {Nat. Photonics}\
       }\textbf {\bibinfo {volume} {6}},\ \bibinfo {pages} {540} (\bibinfo {year}
       {2012})}\BibitemShut {NoStop}%
     \bibitem [{\citenamefont {Milne}\ \emph {et~al.}(2017)\citenamefont {Milne},
       \citenamefont {Schietinger}, \citenamefont {Aiba}, \citenamefont {Alarcon}
       \emph {et~al.}}]{milne2017swissfel}%
       \BibitemOpen
       \bibfield  {author} {\bibinfo {author} {\bibfnamefont {C.}~\bibnamefont
       {Milne}}, \bibinfo {author} {\bibfnamefont {T.}~\bibnamefont {Schietinger}},
       \bibinfo {author} {\bibfnamefont {M.}~\bibnamefont {Aiba}}, \bibinfo {author}
       {\bibfnamefont {A.}~\bibnamefont {Alarcon}},  \emph {et~al.},\ }\href@noop {}
       {\bibfield  {journal} {\bibinfo  {journal} {Appl. Sci.}\ }\textbf {\bibinfo
       {volume} {7}},\ \bibinfo {pages} {720} (\bibinfo {year} {2017})}\BibitemShut
       {NoStop}%
     \bibitem [{\citenamefont {Kang}\ \emph {et~al.}(2017)\citenamefont {Kang},
       \citenamefont {Min}, \citenamefont {Heo} \emph {et~al.}}]{Kang2017}%
       \BibitemOpen
       \bibfield  {author} {\bibinfo {author} {\bibfnamefont {H.~S.}\ \bibnamefont
       {Kang}}, \bibinfo {author} {\bibfnamefont {C.~K.}\ \bibnamefont {Min}},
       \bibinfo {author} {\bibfnamefont {H.}~\bibnamefont {Heo}},  \emph {et~al.},\
       }\href {\doibase 10.1038/s41566-017-0029-8} {\bibfield  {journal} {\bibinfo
       {journal} {Nat. Photonics}\ }\textbf {\bibinfo {volume} {11}},\ \bibinfo
       {pages} {708} (\bibinfo {year} {2017})}\BibitemShut {NoStop}%
     \bibitem [{\citenamefont {Decking}\ \emph {et~al.}(2020)\citenamefont
       {Decking}, \citenamefont {Abeghyan}, \citenamefont {Abramian} \emph
       {et~al.}}]{Decking2020}%
       \BibitemOpen
       \bibfield  {author} {\bibinfo {author} {\bibfnamefont {W.}~\bibnamefont
       {Decking}}, \bibinfo {author} {\bibfnamefont {S.}~\bibnamefont {Abeghyan}},
       \bibinfo {author} {\bibfnamefont {P.}~\bibnamefont {Abramian}},  \emph
       {et~al.},\ }\href {\doibase 10.1038/s41566-020-0607-z} {\bibfield  {journal}
       {\bibinfo  {journal} {Nat. Photonics}\ }\textbf {\bibinfo {volume} {14}},\
       \bibinfo {pages} {391} (\bibinfo {year} {2020})}\BibitemShut {NoStop}%
     \bibitem [{\citenamefont {Pellegrini}\ \emph {et~al.}(2016)\citenamefont
       {Pellegrini}, \citenamefont {Marinelli},\ and\ \citenamefont
       {Reiche}}]{Pellegrini2016}%
       \BibitemOpen
       \bibfield  {author} {\bibinfo {author} {\bibfnamefont {C.}~\bibnamefont
       {Pellegrini}}, \bibinfo {author} {\bibfnamefont {A.}~\bibnamefont
       {Marinelli}}, \ and\ \bibinfo {author} {\bibfnamefont {S.}~\bibnamefont
       {Reiche}},\ }\href {\doibase 10.1103/RevModPhys.88.015006} {\bibfield
       {journal} {\bibinfo  {journal} {Rev. Mod. Phys.}\ }\textbf {\bibinfo {volume}
       {88}} (\bibinfo {year} {2016}),\ 10.1103/RevModPhys.88.015006}\BibitemShut
       {NoStop}%
     \bibitem [{\citenamefont {Feng}\ and\ \citenamefont {Deng}(2018)}]{Feng.2018}%
       \BibitemOpen
       \bibfield  {author} {\bibinfo {author} {\bibfnamefont {C.}~\bibnamefont
       {Feng}}\ and\ \bibinfo {author} {\bibfnamefont {H.~X.}\ \bibnamefont
       {Deng}},\ }\href {\doibase 10.1007/s41365-018-0490-1} {\bibfield  {journal}
       {\bibinfo  {journal} {Nucl. Sci. Tech.}\ }\textbf {\bibinfo {volume} {29}},\
       \bibinfo {pages} {160} (\bibinfo {year} {2018})}\BibitemShut {NoStop}%
     \bibitem [{\citenamefont {Hemsing}\ \emph {et~al.}(2008)\citenamefont
       {Hemsing}, \citenamefont {Gover},\ and\ \citenamefont
       {Rosenzweig}}]{10.1103/physreva.77.063831}%
       \BibitemOpen
       \bibfield  {author} {\bibinfo {author} {\bibfnamefont {E.}~\bibnamefont
       {Hemsing}}, \bibinfo {author} {\bibfnamefont {A.}~\bibnamefont {Gover}}, \
       and\ \bibinfo {author} {\bibfnamefont {J.}~\bibnamefont {Rosenzweig}},\
       }\href {\doibase 10.1103/PhysRevA.77.063831} {\bibfield  {journal} {\bibinfo
       {journal} {Phys. Rev. A - At. Mol. Opt. Phys.}\ }\textbf {\bibinfo {volume}
       {77}},\ \bibinfo {pages} {63831} (\bibinfo {year} {2008})}\BibitemShut
       {NoStop}%
     \bibitem [{\citenamefont {Hemsing}\ \emph {et~al.}(2009)\citenamefont
       {Hemsing}, \citenamefont {Musumeci}, \citenamefont {Reiche}, \citenamefont
       {Tikhoplav} \emph {et~al.}}]{10.1103/physrevlett.102.174801}%
       \BibitemOpen
       \bibfield  {author} {\bibinfo {author} {\bibfnamefont {E.}~\bibnamefont
       {Hemsing}}, \bibinfo {author} {\bibfnamefont {P.}~\bibnamefont {Musumeci}},
       \bibinfo {author} {\bibfnamefont {S.}~\bibnamefont {Reiche}}, \bibinfo
       {author} {\bibfnamefont {R.}~\bibnamefont {Tikhoplav}},  \emph {et~al.},\
       }\href {\doibase 10.1103/PhysRevLett.102.174801} {\bibfield  {journal}
       {\bibinfo  {journal} {Phys. Rev. Lett.}\ }\textbf {\bibinfo {volume} {102}},\
       \bibinfo {pages} {174801} (\bibinfo {year} {2009})}\BibitemShut {NoStop}%
     \bibitem [{\citenamefont {Hemsing}\ \emph {et~al.}(2011)\citenamefont
       {Hemsing}, \citenamefont {Marinelli},\ and\ \citenamefont
       {Rosenzweig}}]{10.1103/physrevlett.106.164803}%
       \BibitemOpen
       \bibfield  {author} {\bibinfo {author} {\bibfnamefont {E.}~\bibnamefont
       {Hemsing}}, \bibinfo {author} {\bibfnamefont {A.}~\bibnamefont {Marinelli}},
       \ and\ \bibinfo {author} {\bibfnamefont {J.~B.}\ \bibnamefont {Rosenzweig}},\
       }\href {\doibase 10.1103/PhysRevLett.106.164803} {\bibfield  {journal}
       {\bibinfo  {journal} {Phys. Rev. Lett.}\ }\textbf {\bibinfo {volume} {106}},\
       \bibinfo {pages} {164803} (\bibinfo {year} {2011})}\BibitemShut {NoStop}%
     \bibitem [{\citenamefont {Hemsing}\ \emph {et~al.}(2013)\citenamefont
       {Hemsing}, \citenamefont {Knyazik}, \citenamefont {Dunning} \emph
       {et~al.}}]{Hemsing2013}%
       \BibitemOpen
       \bibfield  {author} {\bibinfo {author} {\bibfnamefont {E.}~\bibnamefont
       {Hemsing}}, \bibinfo {author} {\bibfnamefont {A.}~\bibnamefont {Knyazik}},
       \bibinfo {author} {\bibfnamefont {M.}~\bibnamefont {Dunning}},  \emph
       {et~al.},\ }\href {\doibase 10.1038/nphys2712} {\bibfield  {journal}
       {\bibinfo  {journal} {Nat. Phys.}\ }\textbf {\bibinfo {volume} {9}},\
       \bibinfo {pages} {549} (\bibinfo {year} {2013})}\BibitemShut {NoStop}%
     \bibitem [{\citenamefont {Ribi{\v{c}}}\ \emph {et~al.}(2014)\citenamefont
       {Ribi{\v{c}}}, \citenamefont {Gauthier},\ and\ \citenamefont {{De
       Ninno}}}]{10.1103/physrevlett.112.203602}%
       \BibitemOpen
       \bibfield  {author} {\bibinfo {author} {\bibfnamefont {P.~R.}\ \bibnamefont
       {Ribi{\v{c}}}}, \bibinfo {author} {\bibfnamefont {D.}~\bibnamefont
       {Gauthier}}, \ and\ \bibinfo {author} {\bibfnamefont {G.}~\bibnamefont {{De
       Ninno}}},\ }\href {\doibase 10.1103/PhysRevLett.112.203602} {\bibfield
       {journal} {\bibinfo  {journal} {Phys. Rev. Lett.}\ }\textbf {\bibinfo
       {volume} {112}},\ \bibinfo {pages} {203602} (\bibinfo {year} {2014})},\
       \Eprint {http://arxiv.org/abs/1312.5837} {arXiv:1312.5837} \BibitemShut
       {NoStop}%
     \bibitem [{\citenamefont {Sasaki}\ and\ \citenamefont
       {McNulty}(2008)}]{10.1103/physrevlett.100.124801}%
       \BibitemOpen
       \bibfield  {author} {\bibinfo {author} {\bibfnamefont {S.}~\bibnamefont
       {Sasaki}}\ and\ \bibinfo {author} {\bibfnamefont {I.}~\bibnamefont
       {McNulty}},\ }\href {\doibase 10.1103/PhysRevLett.100.124801} {\bibfield
       {journal} {\bibinfo  {journal} {Phys. Rev. Lett.}\ }\textbf {\bibinfo
       {volume} {100}},\ \bibinfo {pages} {124801} (\bibinfo {year}
       {2008})}\BibitemShut {NoStop}%
     \bibitem [{\citenamefont {Kim}\ \emph {et~al.}(2008)\citenamefont {Kim},
       \citenamefont {Shvyd'ko},\ and\ \citenamefont {Reiche}}]{Kim2008}%
       \BibitemOpen
       \bibfield  {author} {\bibinfo {author} {\bibfnamefont {K.~J.}\ \bibnamefont
       {Kim}}, \bibinfo {author} {\bibfnamefont {Y.}~\bibnamefont {Shvyd'ko}}, \
       and\ \bibinfo {author} {\bibfnamefont {S.}~\bibnamefont {Reiche}},\ }\href
       {\doibase 10.1103/PhysRevLett.100.244802} {\bibfield  {journal} {\bibinfo
       {journal} {Phys. Rev. Lett.}\ }\textbf {\bibinfo {volume} {100}} (\bibinfo
       {year} {2008}),\ 10.1103/PhysRevLett.100.244802}\BibitemShut {NoStop}%
     \bibitem [{\citenamefont {Dai}\ \emph {et~al.}(2012)\citenamefont {Dai},
       \citenamefont {Deng},\ and\ \citenamefont {Dai}}]{Dai2012}%
       \BibitemOpen
       \bibfield  {author} {\bibinfo {author} {\bibfnamefont {J.}~\bibnamefont
       {Dai}}, \bibinfo {author} {\bibfnamefont {H.}~\bibnamefont {Deng}}, \ and\
       \bibinfo {author} {\bibfnamefont {Z.}~\bibnamefont {Dai}},\ }\href {\doibase
       10.1103/PhysRevLett.108.034802} {\bibfield  {journal} {\bibinfo  {journal}
       {Phys. Rev. Lett.}\ }\textbf {\bibinfo {volume} {108}} (\bibinfo {year}
       {2012}),\ 10.1103/PhysRevLett.108.034802}\BibitemShut {NoStop}%
     \bibitem [{\citenamefont {Li}\ and\ \citenamefont
       {Deng}(2018{\natexlab{a}})}]{Li2018}%
       \BibitemOpen
       \bibfield  {author} {\bibinfo {author} {\bibfnamefont {K.}~\bibnamefont
       {Li}}\ and\ \bibinfo {author} {\bibfnamefont {H.}~\bibnamefont {Deng}},\
       }\href {\doibase 10.1063/1.5037180} {\bibfield  {journal} {\bibinfo
       {journal} {Appl. Phys. Lett.}\ }\textbf {\bibinfo {volume} {113}} (\bibinfo
       {year} {2018}{\natexlab{a}}),\ 10.1063/1.5037180}\BibitemShut {NoStop}%
     \bibitem [{\citenamefont {Best}\ and\ \citenamefont {Faatz}(1990)}]{Best1990}%
       \BibitemOpen
       \bibfield  {author} {\bibinfo {author} {\bibfnamefont {R.~W.}\ \bibnamefont
       {Best}}\ and\ \bibinfo {author} {\bibfnamefont {B.}~\bibnamefont {Faatz}},\
       }\href {\doibase 10.1088/0022-3727/23/11/001} {\bibfield  {journal} {\bibinfo
        {journal} {J. Phys. D. Appl. Phys.}\ }\textbf {\bibinfo {volume} {23}},\
       \bibinfo {pages} {1337} (\bibinfo {year} {1990})}\BibitemShut {NoStop}%
     \bibitem [{\citenamefont {Riyopoulos}\ and\ \citenamefont
       {Tang}(1990)}]{10.1063/1.345634}%
       \BibitemOpen
       \bibfield  {author} {\bibinfo {author} {\bibfnamefont {S.}~\bibnamefont
       {Riyopoulos}}\ and\ \bibinfo {author} {\bibfnamefont {C.~M.}\ \bibnamefont
       {Tang}},\ }\href {\doibase 10.1063/1.345634} {\bibfield  {journal} {\bibinfo
       {journal} {J. Appl. Phys.}\ }\textbf {\bibinfo {volume} {67}},\ \bibinfo
       {pages} {1669} (\bibinfo {year} {1990})}\BibitemShut {NoStop}%
     \bibitem [{\citenamefont {Wu}\ \emph {et~al.}(2019)\citenamefont {Wu} \emph
       {et~al.}}]{wu:fel2019}%
       \BibitemOpen
       \bibfield  {author} {\bibinfo {author} {\bibfnamefont {Y.}~\bibnamefont {Wu}}
       \emph {et~al.},\ }\href@noop {} {\enquote {\bibinfo {title} {{G}enerating
       {O}rbital {A}ngular {M}omentum {B}eams in an {FEL} {O}scillator},}\ }
       (\bibinfo {year} {2019}),\ \bibinfo {note} {presented at FEL2019 in Hamburg,
       Germany, unpublished}\BibitemShut {NoStop}%
     \bibitem [{\citenamefont {Shvydko}\ \emph {et~al.}(2010)\citenamefont
       {Shvydko}, \citenamefont {Stoupin}, \citenamefont {Cunsolo}, \citenamefont
       {Said},\ and\ \citenamefont {Huang}}]{Shvydko2010}%
       \BibitemOpen
       \bibfield  {author} {\bibinfo {author} {\bibfnamefont {Y.~V.}\ \bibnamefont
       {Shvydko}}, \bibinfo {author} {\bibfnamefont {S.}~\bibnamefont {Stoupin}},
       \bibinfo {author} {\bibfnamefont {A.}~\bibnamefont {Cunsolo}}, \bibinfo
       {author} {\bibfnamefont {A.~H.}\ \bibnamefont {Said}}, \ and\ \bibinfo
       {author} {\bibfnamefont {X.}~\bibnamefont {Huang}},\ }\href {\doibase
       10.1038/nphys1506} {\bibfield  {journal} {\bibinfo  {journal} {Nat. Phys.}\
       }\textbf {\bibinfo {volume} {6}},\ \bibinfo {pages} {196} (\bibinfo {year}
       {2010})}\BibitemShut {NoStop}%
     \bibitem [{\citenamefont {Elleaume}\ and\ \citenamefont
       {Deacon}(1984)}]{10.1007/bf00690020}%
       \BibitemOpen
       \bibfield  {author} {\bibinfo {author} {\bibfnamefont {P.}~\bibnamefont
       {Elleaume}}\ and\ \bibinfo {author} {\bibfnamefont {D.~A.}\ \bibnamefont
       {Deacon}},\ }\href {\doibase 10.1007/BF00690020} {\bibfield  {journal}
       {\bibinfo  {journal} {Appl. Phys. B Photophysics Laser Chem.}\ }\textbf
       {\bibinfo {volume} {33}},\ \bibinfo {pages} {9} (\bibinfo {year}
       {1984})}\BibitemShut {NoStop}%
     \bibitem [{\citenamefont {Faatz}\ \emph {et~al.}(1993)\citenamefont {Faatz},
       \citenamefont {Best}, \citenamefont {Oepts},\ and\ \citenamefont {{Van
       Amersfoort}}}]{10.1088/0963-9659/2/3/006}%
       \BibitemOpen
       \bibfield  {author} {\bibinfo {author} {\bibfnamefont {B.}~\bibnamefont
       {Faatz}}, \bibinfo {author} {\bibfnamefont {R.~W.}\ \bibnamefont {Best}},
       \bibinfo {author} {\bibfnamefont {D.}~\bibnamefont {Oepts}}, \ and\ \bibinfo
       {author} {\bibfnamefont {P.~W.}\ \bibnamefont {{Van Amersfoort}}},\ }\href
       {\doibase 10.1088/0963-9659/2/3/006} {\bibfield  {journal} {\bibinfo
       {journal} {Pure Appl. Opt. J. Eur. Opt. Soc. Part A}\ }\textbf {\bibinfo
       {volume} {2}},\ \bibinfo {pages} {195} (\bibinfo {year} {1993})}\BibitemShut
       {NoStop}%
     \bibitem [{\citenamefont {Padgett}\ \emph {et~al.}(2015)\citenamefont
       {Padgett}, \citenamefont {Miatto}, \citenamefont {Lavery}, \citenamefont
       {Zeilinger},\ and\ \citenamefont {Boyd}}]{Padgett2015}%
       \BibitemOpen
       \bibfield  {author} {\bibinfo {author} {\bibfnamefont {M.~J.}\ \bibnamefont
       {Padgett}}, \bibinfo {author} {\bibfnamefont {F.~M.}\ \bibnamefont {Miatto}},
       \bibinfo {author} {\bibfnamefont {M.~P.}\ \bibnamefont {Lavery}}, \bibinfo
       {author} {\bibfnamefont {A.}~\bibnamefont {Zeilinger}}, \ and\ \bibinfo
       {author} {\bibfnamefont {R.~W.}\ \bibnamefont {Boyd}},\ }\href {\doibase
       10.1088/1367-2630/17/2/023011} {\bibfield  {journal} {\bibinfo  {journal}
       {New J. Phys.}\ }\textbf {\bibinfo {volume} {17}} (\bibinfo {year} {2015}),\
       10.1088/1367-2630/17/2/023011},\ \Eprint {http://arxiv.org/abs/1410.8722}
       {arXiv:1410.8722} \BibitemShut {NoStop}%
     \bibitem [{\citenamefont {Szarmes}(2014)}]{Classical2014}%
       \BibitemOpen
       \bibfield  {author} {\bibinfo {author} {\bibfnamefont {E.~B.}\ \bibnamefont
       {Szarmes}},\ }\href {\doibase 10.1088/978-1-6270-5573-4} {\emph {\bibinfo
       {title} {Classical Theory of Free-Electron Lasers}}},\ 2053-2571\ (\bibinfo
       {publisher} {Morgan \& Claypool Publishers},\ \bibinfo {year}
       {2014})\BibitemShut {NoStop}%
     \bibitem [{\citenamefont {Yan}\ and\ \citenamefont {Deng}(2019)}]{Yan2019}%
       \BibitemOpen
       \bibfield  {author} {\bibinfo {author} {\bibfnamefont {J.}~\bibnamefont
       {Yan}}\ and\ \bibinfo {author} {\bibfnamefont {H.}~\bibnamefont {Deng}},\
       }\href {\doibase 10.1103/PhysRevAccelBeams.22.090701} {\bibfield  {journal}
       {\bibinfo  {journal} {Phys. Rev. Accel. Beams}\ }\textbf {\bibinfo {volume}
       {22}} (\bibinfo {year} {2019}),\
       10.1103/PhysRevAccelBeams.22.090701}\BibitemShut {NoStop}%
     \bibitem [{\citenamefont {Li}\ and\ \citenamefont
       {Deng}(2018{\natexlab{b}})}]{Li2018SCLF}%
       \BibitemOpen
       \bibfield  {author} {\bibinfo {author} {\bibfnamefont {K.}~\bibnamefont
       {Li}}\ and\ \bibinfo {author} {\bibfnamefont {H.}~\bibnamefont {Deng}},\
       }\href {\doibase 10.1016/j.nima.2018.03.072} {\bibfield  {journal} {\bibinfo
       {journal} {Nucl. Instruments Methods Phys. Res. Sect. A}\ }\textbf {\bibinfo
       {volume} {895}},\ \bibinfo {pages} {40} (\bibinfo {year}
       {2018}{\natexlab{b}})}\BibitemShut {NoStop}%
     \bibitem [{\citenamefont {Reiche}(1999)}]{Reiche1999}%
       \BibitemOpen
       \bibfield  {author} {\bibinfo {author} {\bibfnamefont {S.}~\bibnamefont
       {Reiche}},\ }\href {\doibase 10.1016/S0168-9002(99)00114-X} {\bibfield
       {journal} {\bibinfo  {journal} {Nucl. Instruments Methods Phys. Res. Sect. A
       Accel. Spectrometers, Detect. Assoc. Equip.}\ }\textbf {\bibinfo {volume}
       {429}},\ \bibinfo {pages} {243} (\bibinfo {year} {1999})}\BibitemShut
       {NoStop}%
     \bibitem [{\citenamefont {{Van Der Slot}}\ \emph {et~al.}(2009)\citenamefont
       {{Van Der Slot}}, \citenamefont {Freund}, \citenamefont {Miner},
       \citenamefont {Benson}, \citenamefont {Shinn},\ and\ \citenamefont
       {Boller}}]{VanDerSlot2009}%
       \BibitemOpen
       \bibfield  {author} {\bibinfo {author} {\bibfnamefont {P.~J.}\ \bibnamefont
       {{Van Der Slot}}}, \bibinfo {author} {\bibfnamefont {H.~P.}\ \bibnamefont
       {Freund}}, \bibinfo {author} {\bibfnamefont {W.~H.}\ \bibnamefont {Miner}},
       \bibinfo {author} {\bibfnamefont {S.~V.}\ \bibnamefont {Benson}}, \bibinfo
       {author} {\bibfnamefont {M.}~\bibnamefont {Shinn}}, \ and\ \bibinfo {author}
       {\bibfnamefont {K.~J.}\ \bibnamefont {Boller}},\ }\href {\doibase
       10.1103/PhysRevLett.102.244802} {\bibfield  {journal} {\bibinfo  {journal}
       {Phys. Rev. Lett.}\ }\textbf {\bibinfo {volume} {102}} (\bibinfo {year}
       {2009}),\ 10.1103/PhysRevLett.102.244802}\BibitemShut {NoStop}%
     \bibitem [{\citenamefont {Huang}\ \emph {et~al.}(2019)\citenamefont {Huang},
       \citenamefont {Li},\ and\ \citenamefont {Deng}}]{Huang2019}%
       \BibitemOpen
       \bibfield  {author} {\bibinfo {author} {\bibfnamefont {N.~S.}\ \bibnamefont
       {Huang}}, \bibinfo {author} {\bibfnamefont {K.}~\bibnamefont {Li}}, \ and\
       \bibinfo {author} {\bibfnamefont {H.~X.}\ \bibnamefont {Deng}},\ }\href
       {\doibase 10.1007/s41365-019-0559-5} {\bibfield  {journal} {\bibinfo
       {journal} {Nucl. Sci. Tech.}\ }\textbf {\bibinfo {volume} {30}} (\bibinfo
       {year} {2019}),\ 10.1007/s41365-019-0559-5}\BibitemShut {NoStop}%
     \end{thebibliography}
%

\end{document}